\documentclass{PoS}

\usepackage[pdftex]{graphicx}
\usepackage{bbm}
\usepackage{verbatim}
\usepackage{psfrag}
\usepackage{slashed}

\usepackage{amsmath}
\usepackage{amssymb}
\usepackage{marvosym}
\usepackage{wasysym}

\def \Ns {{N_{\sigma}}}
\def \Nt {{N_{\tau}}}
\def \at {{a_{\tau}}}
\def \Nc {{N_{c}}}

\def \hmu {{\hat{\mu}}}

\newcommand{\lr}[1]{\left( #1 \right)}

\newcommand{\nn} {\nonumber\\}
\newcommand{\beqn} {\begin{equation}}
\newcommand{\eqn} {\end{equation}}

\def \beq{\begin{equation}}
\def \eeq{\end{equation}}
\def \bea{\begin{eqnarray}}
\def \eea{\end{eqnarray}}

\def \bet0{\beta_0}
\def \bet1{\beta_1}
\def \simgt{\,\rlap{\lower 7.5 pt\hbox{$\mathchar \sim$}}\raise 3 pt \hbox{$>$}\,}
\def \simlt{\,\rlap{\lower 7.5 pt\hbox{$\mathchar \sim$}}\raise 3 pt \hbox{$<$}\,}

\def\lsim{\raise0.3ex\hbox{$<$\kern-0.75em\raise-1.1ex\hbox{$\sim$}}}
\def\gsim{\raise0.3ex\hbox{$>$\kern-0.75em\raise-1.1ex\hbox{$\sim$}}}

\def \Zcal{\mathcal{Z}}

\title{
Continuous Time Monte Carlo for Lattice
QCD\\ in the Strong Coupling Limit}

\ShortTitle{Continuous Time Monte Carlo for Lattice QCD in the Strong Coupling Limit}

\author{\speaker{Wolfgang Unger}\\
        Institut f\"ur theoretische Physik, ETH Z\"urich, CH-8093, Switzerland\\
        E-mail: \email{ungerw@phys.ethz.ch}
}

\author{Philippe de Forcrand\\
        Institut f\"ur theoretische Physik, ETH Z\"urich, CH-8093, Switzerland\\
        CERN, Physics Department, TH Unit, CH-1211 Geneva 23, Switzerland\\
        E-mail: \email{forcrand@phys.ethz.ch}
}

\abstract{
\vspace*{-10.5cm}
\begin{flushright}
\texttt{\footnotesize CERN-PH-TH/2011-273}\\
\end{flushright}
\vspace*{9.5cm}
We present results for lattice QCD with staggered fermions in the limit of infinite gauge coupling, obtained from a worm-type Monte Carlo algorithm on a discrete spatial lattice
but with continuous Euclidean time. This is achieved by sending both the anisotropy parameter $\gamma^2\simeq a/\at$ and the number of time-slices $N_\tau$ to infinity,
keeping the ratio $\gamma^2/N_\tau \simeq aT$ fixed. 
In this limit, ambiguities arising from the anisotropy parameter $\gamma$ are eliminated
and discretization errors usually introduced by a finite temporal lattice extent $\Nt$ are absent.
The obvious gain is that no continuum extrapolation $N_\tau \rightarrow \infty$ has to be carried out.
Moreover, the algorithm is faster and the sign problem disappears completely. 

As a first application, we determine the phase diagram as a function of temperature and real and imaginary baryon chemical potential.  
We compare our computations with those on lattices with discrete Euclidean time.
Discretization errors due to finite $\Nt$ in previous studies turn out to be large at low temperatures.
}
\FullConference{ The XXIX International Symposium on Lattice Field Theory - Lattice 2011\\
July 10-16, 2011\\
Squaw Valley, Lake Tahoe, California}

\begin{document}

\vspace{-1mm}
\section{Introduction}
\vspace{-2mm}

The determination of the QCD phase diagram, in particular the location of the critical point, is an important, long standing problem, requiring non-perturbative methods.
In lattice QCD, several approaches have been developed to investigate the phase transition from the hadronic matter to the quark gluon plasma,
but all of them are limited to $\mu_B/T \lesssim 1$, with $\mu_B$ the baryon chemical potential \cite{Forcrand2009}.
The reason for this is the notorious sign problem, which arises because the fermion determinant for finite $\mu_B$ becomes complex, and importance sampling is no longer applicable. 
In QCD, the sign problem is severe. The relative fluctuations of the complex phase factor grow exponentially with the lattice volume.
However, in the strong coupling limit of lattice QCD (SC-QCD) discussed below, 
the order of integration is reversed: the gauge links are integrated out first, and the partition function is expressed as a gas of hadron world lines. There is no 
fermion determinant, and the sign problem is much milder. This allows us to obtain the full $(\mu_B,T)$ phase diagram.

\vspace{-1mm}
\section{Strong Coupling Lattice QCD in the Continuous Time Formulation}
\vspace{-2mm}


In SC-QCD, the gauge coupling is sent to infinity and hence the coefficient of the plaquette term $\beta=6/g^2$ is sent to zero. Thus, the Yang Mills part $F_{\mu\nu} F_{\mu\nu}$ of the action is absent.
Then, the gauge fields in the covariant derivative can be integrated out analytically. 
However, as a consequence of the strong coupling limit, the lattice spacing $a$  becomes very large, and no continuum limit can be considered.
The degrees of freedom in SC-QCD live on a crystal.
We study the SC limit for one flavor of staggered fermions.\footnote{
In the continuum limit of staggered fermions, one flavor represents four degenerate tastes. In the strong coupling limit,
one simply has one flavor of spinless fermions on a crystal.
}
The action is given by the fermionic part only:
\begin{eqnarray}
S[U,\chi,\bar{\chi}]=am_q\sum_x\bar{\chi}(x) \chi(x) 
&+& \frac{\gamma}{2}\sum_x \eta_0(x) \left[ \bar{\chi}(x) e^{\at\mu} U_0(x)\chi(x+\hat{0})- \bar{\chi}(x+\hat{0}) e^{-\at\mu} U_0^\dagger(x)\chi(x)\right]\nn
&+& \frac{1}{2}\sum_x  \sum_i^d \eta_i(x) \left[ \bar{\chi}(x) U_i(x)\chi(x+\hat{i})- \bar{\chi}(x+\hat{i}) U_i^\dagger(x)\chi(x)\right]
\label{SCQCDPF}
\end{eqnarray}
with $m_q$ the quark mass and $\mu=\frac{1}{3}\mu_B$ the quark chemical potential. The anisotropy parameter $\gamma$ modifies the Dirac coupling of the temporal part. It will be discussed in detail below. 

Following the procedure discussed in \cite{Rossi1984}, the gauge link integration for gauge group SU($\Nc$) can be performed analytically, as the integration factorizes in Eq.~(\ref{SCQCDPF}):
\begin{eqnarray}
\Zcal
&=&\int \prod_x d\bar{\chi} d\chi e^{2am_q M(x)}\prod_{\mu} \text{\huge$\{$}  \sum_{k_\mu(x)=0}^{\Nc}\frac{(\Nc - k_\mu(x))!}{\Nc! k_\mu(x)!} 
\lr{\lr{\eta_\hmu(x)\gamma^{\delta_{0,\mu}}}^2 M(x)M(x+\hmu)}^{k_\mu(x)}  \nonumber\\ 
&&+ \kappa \lr{\rho(x,y)^{\Nc}\bar{B}(x)B(x+\hmu)+(-\rho(y,x))^{\Nc}\bar{B}(x+\hmu)B(x)}\text{\huge $\}$}  \label{PFGI}\\
M(x)&\equiv&\bar{\chi}(x)\chi(x), \qquad B(x)\equiv\frac{1}{\Nc!}\epsilon_{i_1\ldots i_{\Nc}}\chi_{i_1}(x)\ldots\chi_{i_{\Nc}}(x)
\end{eqnarray}
The new degrees of freedom are mesons $M(x)$ and baryons $B(x)$:
The first part of the expression (\ref{PFGI}) describes the mesonic sector (where $k_\mu(x)$ counts the number of meson hoppings),
and the second part describes the baryonic sector, which involves a $\mu$-dependent weight
$\rho(x,y)$.\footnote{Note that baryons transform under gauge transformation $\Omega\in U(3)$ as $B(x) \rightarrow B(x) \det\Omega$,
hence they are not U(3) gauge invariant. U(3) describes a purely mesonic system ($\kappa=0$ in Eq.~(\ref{PFGI})), while SU(3) contains baryons ($\kappa =1$).}
After performing the Grassmann integrals analytically, the strong coupling partition function is:
\begin{eqnarray}
\Zcal(m_q,\mu_q)= \sum_{\{k,n,l\}}\prod_{b=(x,\hat{\mu})}\frac{(\Nc-k_b)!}{\Nc!k_b!}
\gamma^{2k_b\delta_{\hat{0}\hat{\mu}}}\prod_{x}\frac{\Nc!}{n_x!}(2am_q)^{n_x} \prod_l w(l)
\label{SCPFF}\\
\text{with the constraint:}\qquad n_x+\sum_{\hat{\mu}=\pm\hat{0},\ldots \pm \hat{3}} k_{\hat{\mu}}(x)=\Nc, \quad 
\forall x\in V_M
\label{GC}
\end{eqnarray}
This system, obtained by an exact rewriting with no approximation other than the strong coupling limit involved, can be described by 
confined, colorless, discrete degrees of freedom:
\begin{itemize} 
\item Mesonic degrees of freedom: 
monomers $n_x \in \{0,\ldots \Nc\}$ and dimers $k_{\hat{\mu}}(x)\in \{0,\ldots \Nc\}$ which are non-oriented meson hoppings. 
They obey the Grassmann constraint Eq.~(\ref{GC}) on mesonic sites $V_M\subset \Ns^3\times\Nt$.\vspace{-2mm}
\item Baryonic degrees of freedom: they form oriented, self-avoiding loops $l$ with 
weight $w(l)$ involving the chemical potential,
the sign $\sigma(l)=\pm 1$ and winding number $r(l)$ which depend on the geometry of the loops. Baryonic sites are not touched by mesons: $V_B \dot{\cup} V_M =\Ns^3\times\Nt$.
\end{itemize}
Here, we consider the chiral limit, $m_q=0$. Then, from Eq.~(\ref{SCPFF}), monomers are absent: $n_x=0$.\\

\newcommand{\Ord} {\mathcal{O}}

\begin{figure}[tbp!]
\vspace{-5mm}
\includegraphics[width=0.47\textwidth]{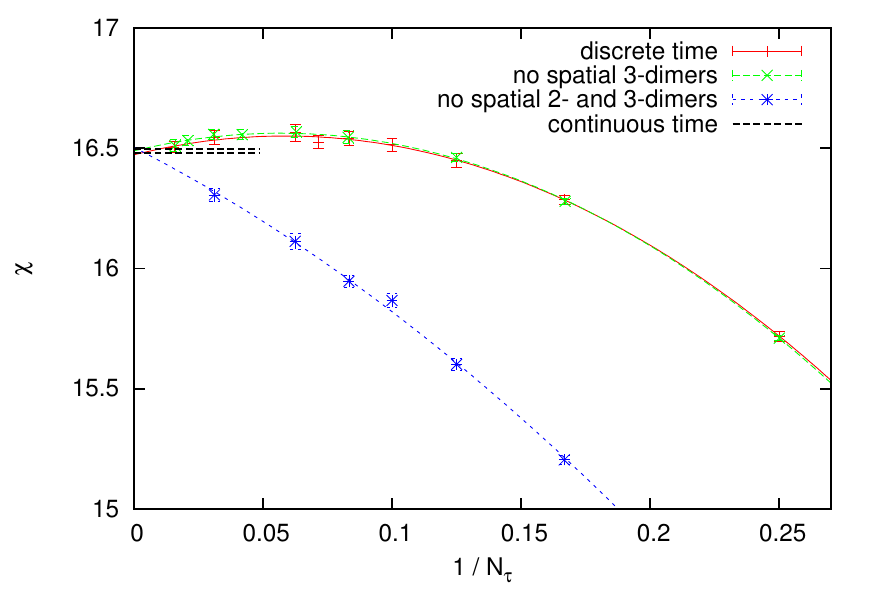}
\includegraphics[width=0.47\textwidth]{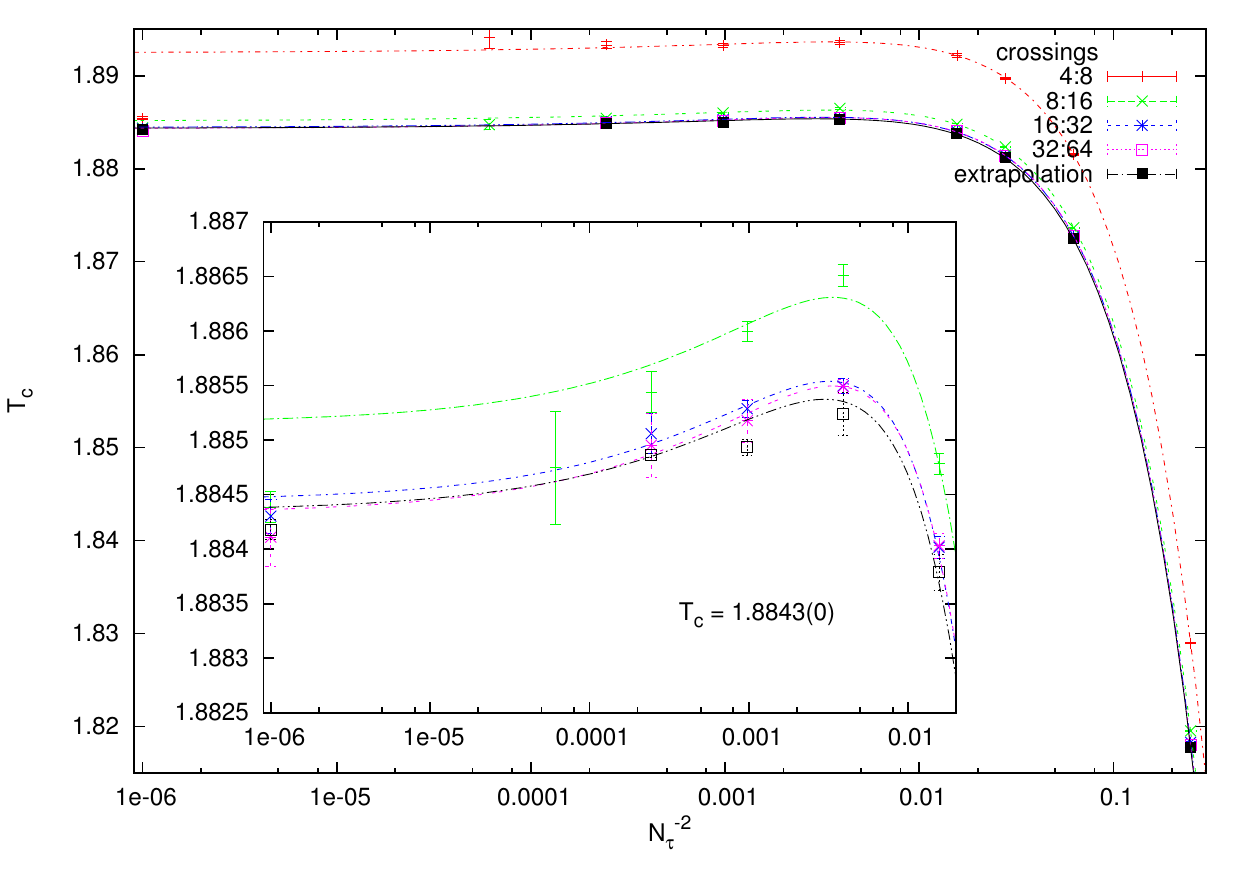}
\begin{minipage}{0.47\textwidth}
\includegraphics[width=\textwidth]{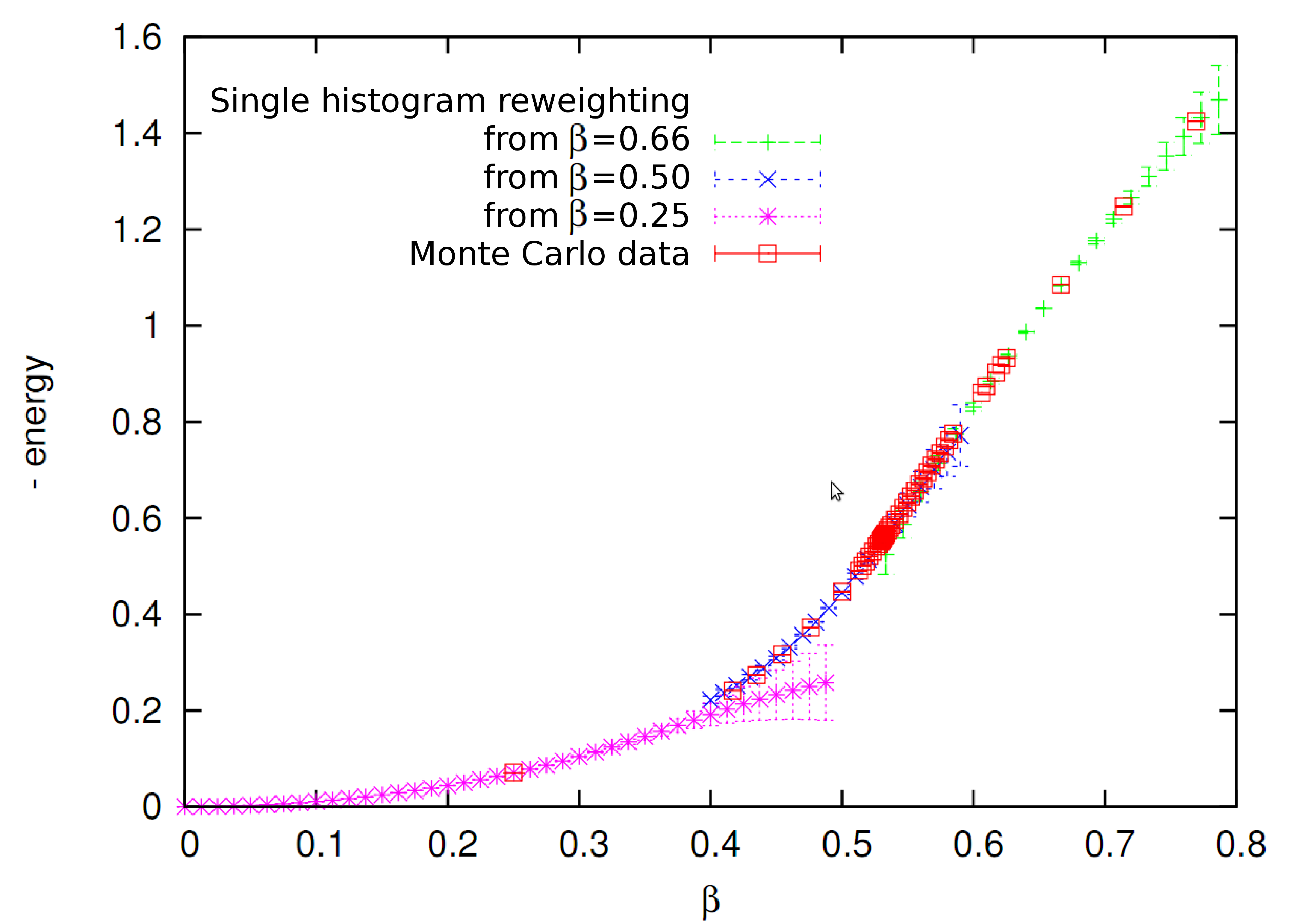}
\end{minipage}\quad
\begin{minipage}{0.5\textwidth}
\vspace{-5mm}
\caption{\emph{Top left:} 
continuous-time limit of the chiral susceptibility $\chi$ at $\gamma^2/\Nt=1.8$ 
in the purely mesonic system U(3), exhibiting typical non-monotonic behavior in $1/\Nt$. 
Note that the suppression of double and/or triple spatial dimers has no effect on the CT limit.
\emph{Top right:} Non-monotonic continuum extrapolation for $aT_c$ and comparison with the continuous time limit.
\emph{Bottom:} Reweighting of the energy based on the partition function Eq.~(2.12).}
\label{NONMONOTONIC}
\end{minipage}
\vspace{-5mm}
\end{figure}

In Eq.~(\ref{SCQCDPF}) we have introduced an anisotropy $\gamma$ in the Dirac couplings. This complication is necessary because the chiral restoration temperature
is given by roughly $a T\simeq1.5$, and on an isotropic lattice with $aT=1/\Nt$ we could not
address the physics of interest. Moreover, with the plaquette term being zero, varying $\gamma$ is the only way to vary the temperature continuously.
The temperature, given by the inverse of the lattice extent in the temporal direction, is thus
\begin{equation}
 T=\frac{f(\gamma)}{a \Nt} \qquad \text{with} \qquad f(\gamma)=a/\at.
\end{equation}
However, the functional dependence $f(\gamma)$ of the ratio of the spatial and temporal lattice spacings on $\gamma$ is not known.
Naive inspection of the derivatives in Eq.~(\ref{SCQCDPF}) would indicate $f(\gamma)=\gamma$, but this only holds at weak coupling.
The mean field approximation of SC-QCD for SU($\Nc$) gauge group  based on $1/d$-expansion with $d$ the spatial dimension \cite{Bilic1992a} yields 
for the critical anisotropy
\begin{equation}
 \gamma_c^2=\Nt \frac{d (\Nc+1)(\Nc+2)}{6(\Nc+3)},
\end{equation}
suggesting that $aT_c=\frac{\gamma_c^2}{\Nt}$ is the sensible, $\Nt$-independent identification in leading order in $1/d$. 
This is also confirmed numerically: In Fig.~\ref{NONMONOTONIC} (top left) we show the variation with $\Nt$ of the chiral susceptibility
close to the U(3) chiral transition,
and in Fig.~\ref{NONMONOTONIC} (top right) we show the variation with $\Nt$ of the U(3) transition temperature defined as $\gamma_c^2/\Nt$. Both quantities approach continuous time (CT) limits, but not monotonically.
Corrections to the CT limit, e.g.~for the critical temperature $T_c$, can be parameterized as
\begin{eqnarray}
aT_{c}(\Nt)&=&aT_{c}^{CT} + B/\Nt + C/\Nt^2 + \Ord(\Nt^{-3})
\end{eqnarray}
where $B$ and $C$ have opposite signs.
To circumvent this difficult extrapolation problem, we have developed an algorithm which samples directly the CT partition function in the limit
\begin{equation}
\Nt\rightarrow \infty, \qquad \gamma \rightarrow \infty, \qquad \gamma^2/\Nt\equiv aT \qquad {\rm fixed.}
\end{equation}
Hence we are left with only one parameter setting the thermal properties, and all discretization errors introduced by a finite $\Nt$ are removed.
Additionally, an algorithm operating in this limit has several advantages: 
There is no need to perform the continuum extrapolation $\Nt \rightarrow \infty$, which 
allows to estimate critical temperatures more precisely, with a faster algorithm. 
And in the baryonic sector of the partition function great simplifications occur: Baryons become static in the CT limit, hence the sign problem is completely absent.

We now explain the partition function we have used for Monte Carlo in the CT limit: after factorizing  Eq.~(\ref{SCPFF}) into spatial and temporal parts, 
the spatial part simplifies greatly when taking into account which configurations
are suppressed with powers of $\gamma^{-1}$:
\begin{eqnarray}
\mathcal{Z}(\gamma, \Nt) 
&\underset{\gamma \rightarrow \infty}{=}& \gamma^{\Nc V} \left(\sum_{\{k\}}\prod_{b_\sigma=(x,\hat{i})}\frac{1}{\Nc}\gamma^{-2k_{b_\sigma}}\prod_{b_\tau=(x,\hat{0})}\frac{(\Nc-k_{b_\tau})!}{\Nc!k_{b_\tau}!}\right)
\left(\prod_{x\in V_B}e^{-3\sigma(x) \mu \Nt/\gamma^{2}}\right).
\label{PARF}
\end{eqnarray}
The identity holds only approximately if $\gamma$ is finite, but becomes exact as $\gamma \rightarrow \infty$ 
because spatial dimers with multiplicity $k_i(x)>1$ are suppressed by powers of $\gamma$ and hence disappear when $\gamma\rightarrow \infty$. 
This is illustrated in Fig.~\ref{absorptionemission} (right): as the temporal lattice spacing $\at\simeq a/\gamma^2 \rightarrow 0$,
multiple spatial dimers become resolved into single dimers. 
The overall number of spatial dimers remains finite in the CT limit, as the sum over ${\cal O}(\gamma^2)$ sites compensates the $1/\gamma^2$ suppression. 
Temporal dimers can be arranged in chains of alternating $k$-dimers and $(\Nc-k)$-dimers. In particular, for $\Nc=3$, we denote 3-0-chains as dashed lines, and 2-1-chains
as solid lines (see Fig.~\ref{absorptionemission}). 

The crucial observation is that the weight of these chains in the partition function is independent of their length,
because the weight of each $k$-dimer is the inverse of that of the $(\Nc-k)$-dimer.
Hence, the weight of a configuration will only depend on the kind and number of vertices at which spatial hoppings are attached to solid/dashed lines,
not on their positions.

For SC-QCD with $\Nc=3$, there are two kinds of vertices, ``L''-vertices of weight $v_L=\gamma^{-1}$, where dashed and solid lines join,
and ``T''-vertices of weight $v_T=2\gamma^{-1}/\sqrt{3}$, where a solid line emits/absorbs a spatial dimer.
The partition function can now be written in terms of these vertices:
\begin{eqnarray}
\mathcal{Z}(\gamma, \Nt) &\underset{\gamma \rightarrow \infty}{=} & \sum_{\mathcal{G}} e^{3\mu B \Nt/\gamma^2} \prod_{x\in V_M} \left(\frac{\hat{v}_L}{\gamma}\right)^{n_L(x)} \left(\frac{\hat{v}_T}{\gamma}\right)^{n_T(x)}, \qquad \hat{v}_L=1, \quad \hat{v}_T=2/\sqrt{3}.
\label{PARF}
\end{eqnarray}
The exponents $n_L(x)$ and $n_T(x)$ in Eq.~(\ref{PARF}) denote the number of L-vertices and T-vertices at spatial site $x\in V_M$.
In contrast to meson hoppings, spatial baryon hoppings are completely suppressed in the CT limit by a factor $\gamma^2/\gamma^3=\gamma^{-1}$. 
Hence, baryons are static in continuous time and the sign problem has completely vanished!
In the limit $\Nt\rightarrow \infty$ the CT partition function can be written simply in terms of $\beta=\Nt/\gamma^2$, up to some normalization constant:
\begin{equation}
\mathcal{Z}_{\rm CT}(\beta)=\sum_{k\in 2\mathbb{N}}(\beta/2)^{k}\sum_{\mathcal{G}'\in \Gamma_k} e^{3\mu B \beta} \hat{v}_T^{N_T}
\quad\text{with}\quad
k=\sum_{b=(x,\hat{i})} k_b, \qquad N_T=\sum_x n_T(x)
\label{PARFCT}
\end{equation}
where $\Gamma_k$ is the set of all configurations $\mathcal{G}'$ equivalent up to time shifts of the vertices and with a total number $k$ of spatial hoppings, and $B$ is the baryon number.
In Fig. \ref{NONMONOTONIC} (bottom) we illustrate the single histogram reweighting of the energy in $\beta$ based on this partition function.
An important property of the above partition function Eq.~(\ref{PARFCT}) is that spatial dimers are distributed uniformly in time.
The lengths $\Delta\beta$ of dashed or solid time intervals (which are related to the number of L- and T-vertices) are then, according to a Poisson process, exponentially distributed:
\begin{equation}
P(\Delta\beta)\sim \exp(-\lambda \Delta\beta),\qquad \Delta\beta \in [0,\beta=1/aT]
\end{equation}
\begin{equation}
\lambda=d_M(x)/4, \qquad d_M(x)=2d-\sum_{\mu} n_B(x\pm\hmu)
\end{equation}
with $\lambda$ the ``decay constant'' for spatial dimer emissions. 
The presence of baryons results in $\lambda$ being space dependent, with $d_M(x)$ the number of mesonic neighbors at a given coordinate $x$. 
Non-trivial meson correlations arise from the entropy of the various configurations. 
Likewise, baryonic interactions beyond the original hard core repulsion are due to the modification they induce on the meson bath, and thus also arise from entropy.

\vspace{-1mm}
\section{Continuous Time Worm Algorithm}
\vspace{-2mm}

\begin{figure}[btp!]
\vspace{-5mm}
\begin{minipage}{0.45\textwidth}
\includegraphics[width=\textwidth]{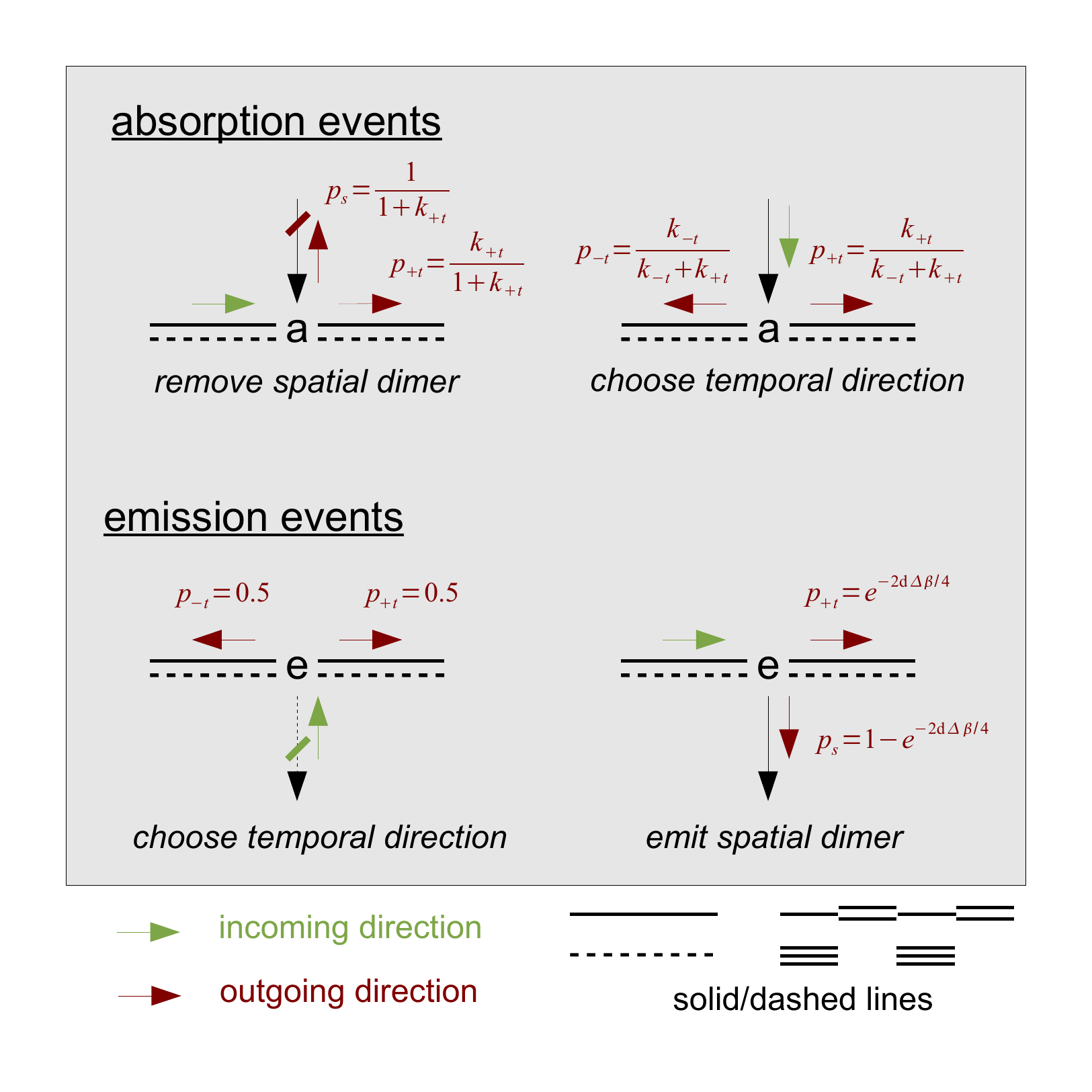}
\end{minipage}
\begin{minipage}{0.5\textwidth}
\vspace{-5mm}
\includegraphics[width=0.52\textwidth]{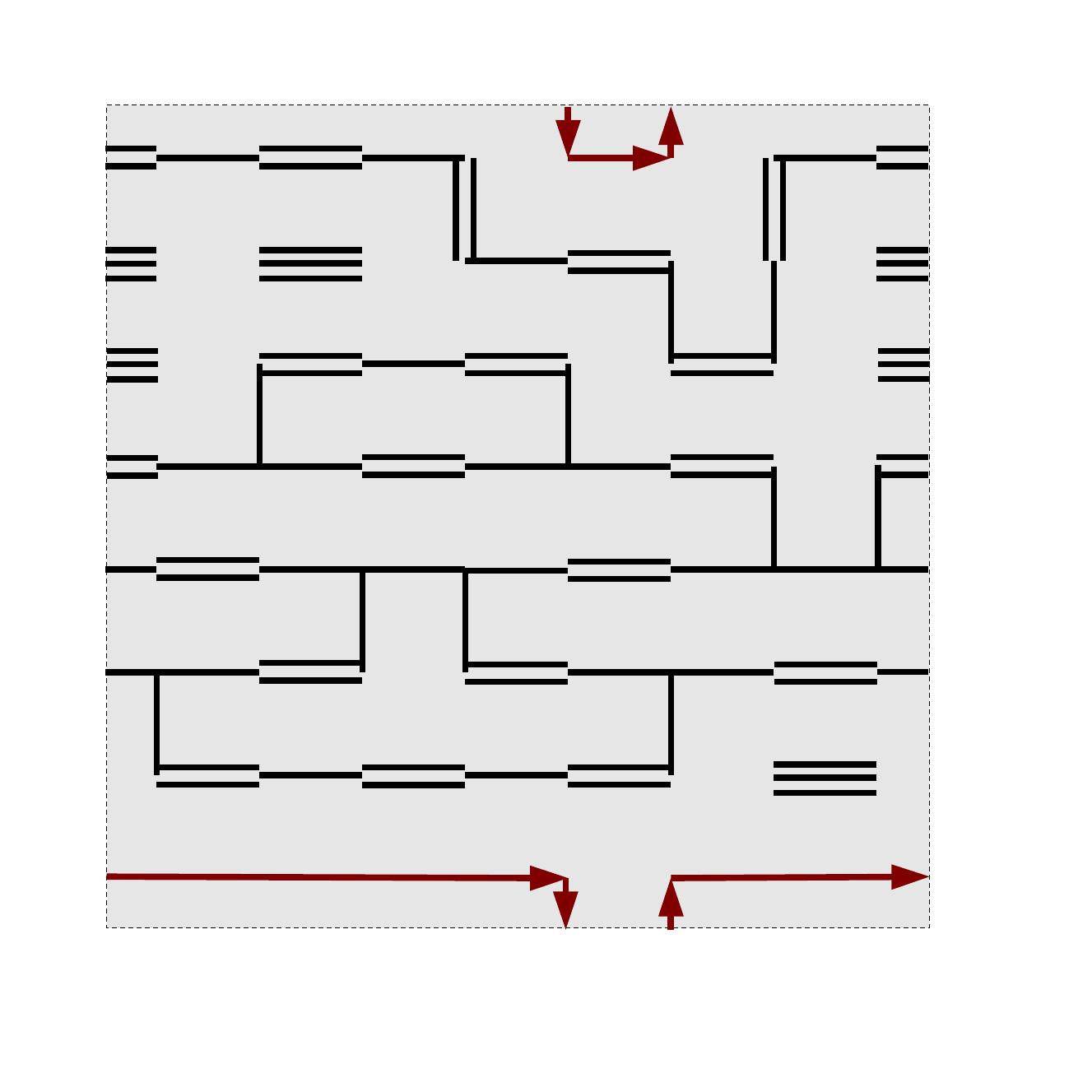}\hspace{-6mm}
\includegraphics[width=0.52\textwidth]{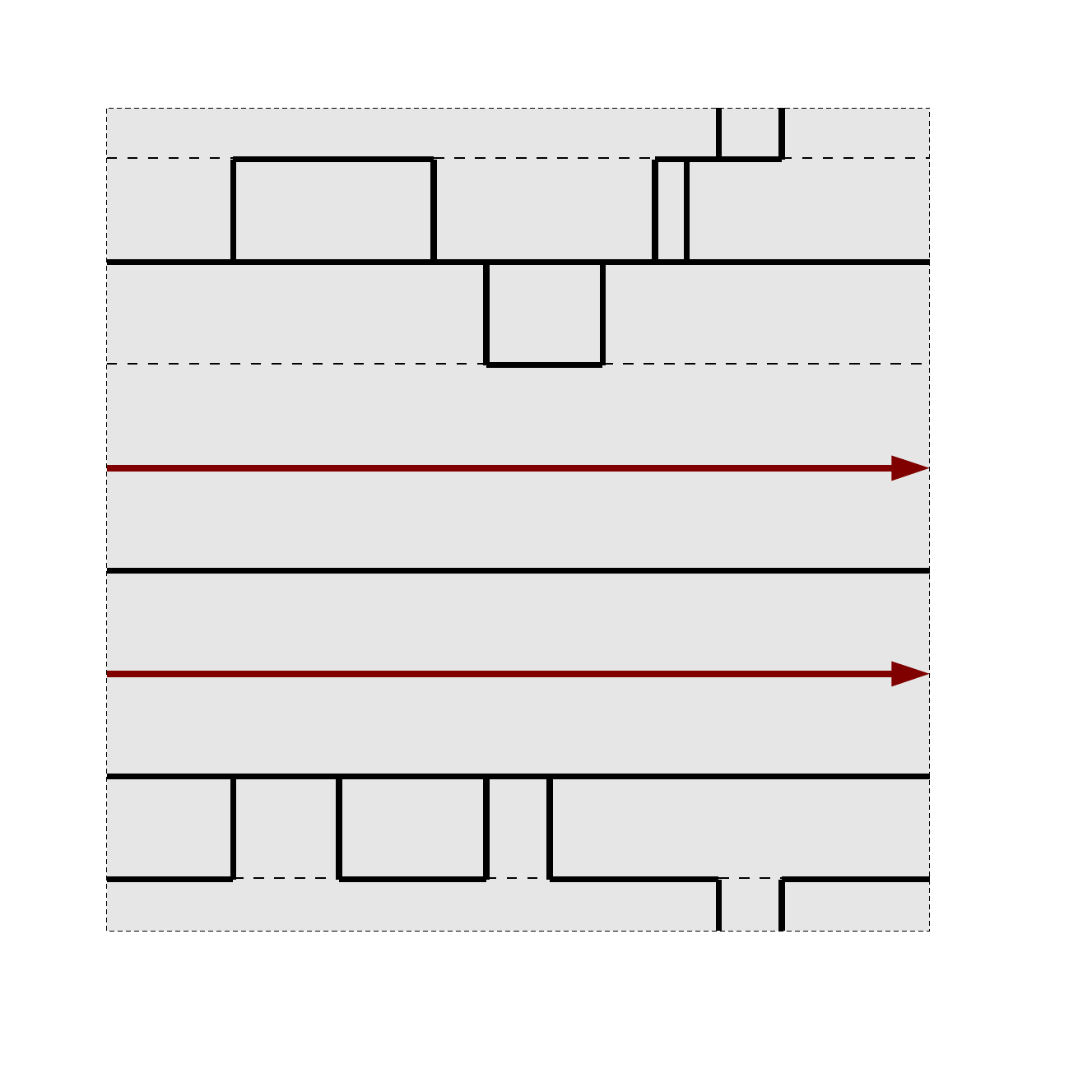}
\vspace{-5mm}
\caption{
\label{absorptionemission}
\emph{Left:} updating rules for the continuous time algorithm. 
 \emph{Top:} illustrative 2-dim.~configurations (time flows to the right), in discrete time ({\em left}) and continuous time ({\em right}). 
Note how the latter lacks multiple spatial dimers and has only static baryons lines.}
\end{minipage}
\vspace{-7mm}\\
\end{figure}

Continuous time (CT) algorithms are now widely used in quantum Monte Carlo (see e.g. \cite{Beard1996, Gull2010}), but to our knowledge have not yet been applied to quantum field theories.
Special difficulties associated with the local gauge symmetry are absent in our case, since gauge fields have been analytically integrated out.
The CT algorithm used here is a Worm-type algorithm, similar to the directed path algorithm introduced for SC-QCD in \cite{Adams:2003cca}. 
In the latter, sites are partitioned into ``active'' and ``passive'' depending on their parity with respect to the worm tail position. Here, we similarly
have ``emission'' and ``absorption'' sites, giving the spatial dimers a consistent orientation.
In Fig.~\ref{absorptionemission} we outline the updating rules of the continuous time worm, and also show a typical 2-dimensional configuration in continuous time. 

In contrast to simulations at finite $\Nt$, there is no need for a baryonic worm update. A simple heatbath update on static lines with no spatial dimers attached is sufficient:
positively (negatively) oriented baryons are (dis)favored by a factor $\exp(\pm 3\mu/T)$ over static meson lines, and their weight is always positive.
We found it very useful to resum baryonic and static mesonic degrees of freedom into so-called static polymers similarly to the MDP algorithm \cite{Karsch1989}: 
first, this allows to extend simulations to arbitrary imaginary chemical potential.
Second, this enables us to adapt the Wang-Landau method for determining the nuclear transition at low 
temperature very accurately. This technique reduces the uncertainty on the first order line $\mu_c(T)$ greatly.
The polymer formulation we have used resums static mesons, baryons and anti-baryons. A configuration with $P$ static polymers has weight\footnote{
Note that there are 4 kinds of static mesons: dashed lines and solid lines of even and odd parity.}
$w(P)=(4+2 \cosh(3\mu/T))^P$.
In our simulation, we only keep track of the polymer number $P=D_0+B^++B^-$, which is the sum of the numbers
$D_0$, $B^+$, and $B^-$ of static mesons, baryons and anti-baryons, respectively.
We calculate the baryon number $B=B^+-B^-$ from $P$ via a trinomial distribution. In the Wang-Landau method, 
we hence obtain at fixed $T$ the density of states for baryon number $B$, $g(B,T)$, from the density of states for static polymer number $P$, $g(P,T)$.

\vspace{-1mm}
\section{Results on the SC-QCD Phase Diagram}
\vspace{-2mm}

\begin{figure}[tbp!]
\vspace{-5mm}
\includegraphics[width=0.47\textwidth]{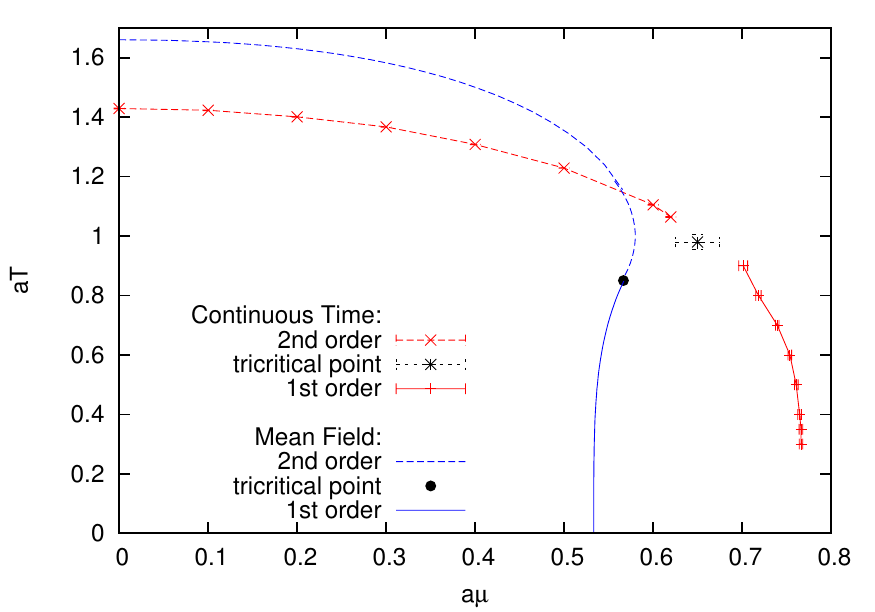}\quad
\includegraphics[width=0.47\textwidth]{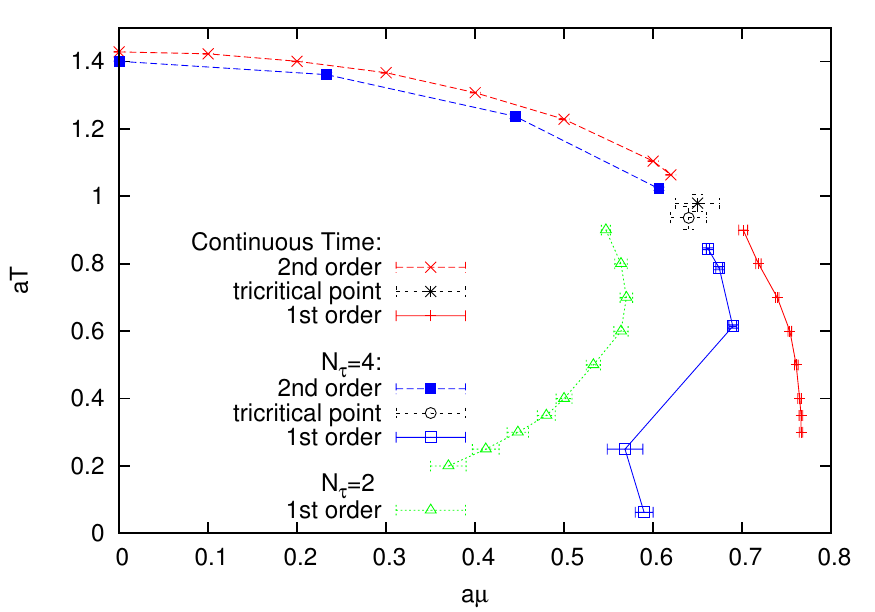}\\
\begin{minipage}{0.47\textwidth}
\includegraphics[width=\textwidth]{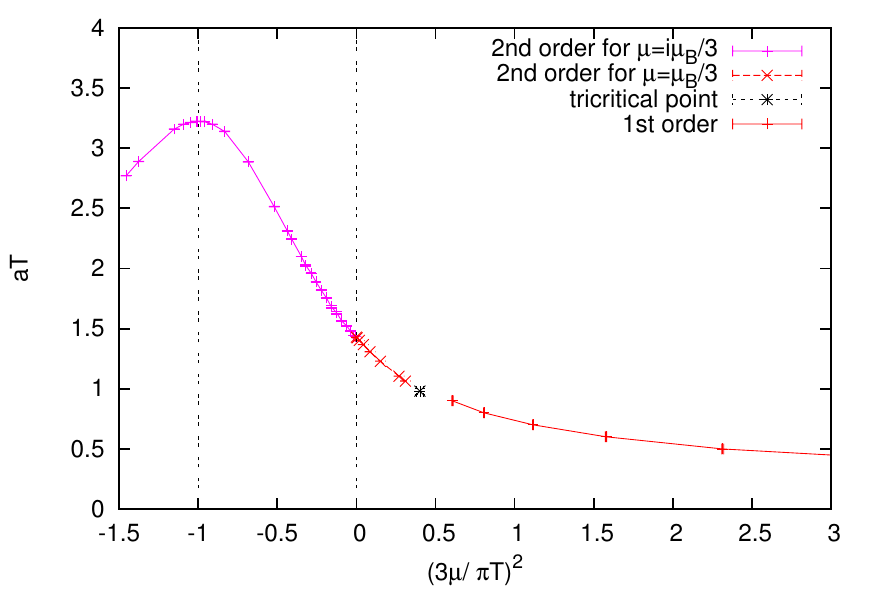}
\end{minipage}\quad
\begin{minipage}{0.49\textwidth}
\vspace{-4mm}
\caption{\emph{Top left:} Comparison of SC-QCD phase diagram obtained from mean field~\cite{Nishida2004} and in continuous time. The former predicts 
a re-entrance which is not present in CT Monte Carlo.
\emph{Top right:} The SC-QCD phase diagram obtained with $\Nt=2, 4$ using $aT=\gamma^2/\Nt$, $a\mu=\gamma^2(\at \mu)$ \cite{Forcrand2010} and in continuous time.
\emph{Bottom:} Extension of the phase diagram including imaginary chemical potential, with no signal for a Roberge Weiss transition (x=-1) at high $T$.}
\label{PhaseDiag}
\end{minipage}
\vspace{-5mm}
\end{figure}

In SC-QCD at low temperature, chiral symmetry, i.~e.~the $U_A(1)$ symmetry of the one-flavor staggered action, 
is spontaneously broken according to $U_L(1) \times U_R(1) \rightarrow U_V(1)$ and becomes restored at some critical temperature $T_c(\mu)$.
%
%
Our new results for the phase boundary $T_c(\mu)$ in the chiral limit $m_q=0$ eliminate systematic errors affecting previous findings based on mean field approximations \cite{Nishida2004} 
or Monte Carlo for fixed $\Nt$ \cite{Forcrand2010}.
As previously found, the phase transition is second order at small $\mu$, and first order at low $T$. The tricritical point is located close to its earlier
$\Nt=4$ estimate \cite{Forcrand2010}, if one uses $a/\at=\gamma^2$, i.e.~$aT=\gamma^2/\Nt$, $a\mu=\gamma^2(\at \mu)$.
However, the re-entrance predicted by mean field analysis \cite{Nishida2004} (which fixes $\gamma=1$ and varies $\Nt\in \mathbb{R}$) and seen in earlier Monte Carlo studies
\cite{Forcrand2010,Unger2011} is absent, see Fig.~\ref{PhaseDiag} (top left). As illustrated in Fig.~\ref{PhaseDiag} (right) the 
discretization error from $\Nt=4$ or 2 becomes very large at low $T$, and mimics re-entrance.
We have also considered an imaginary chemical potential. As shown in Fig.~\ref{PhaseDiag} (bottom), the phase diagram displays Z(3) periodicity, but the imaginary $\mu$
transition remains second-order and there is no Roberge-Weiss transition at high temperature. This is natural: at high $T$ the partition function Eq.~(\ref{PARFCT}) is dominated by the $k=0$ term
which is analytic in $\mu$.

In summary, the continuous time formalism allows for a final, unambiguous determination of the strong coupling phase diagram, and is suitable for further extensions
like a second quark flavor. 

\vspace{-1mm}
\section{Acknowledgments}
\vspace{-2mm}

The computations have been carried out on the Brutus cluster at the ETH Z\"urich. 
We thank S.~Chandrasekharan, M.~Fromm  and O.~Philipsen for useful discussions.
This work is supported by the Swiss National Science Foundation under grant 200020-122117.
\vspace{-3mm}\\

\end{document}